# Secure SAML validation to prevent XML signature wrapping attacks

*Pawel Krawczyk, pawel.krawczyk@owasp.org,*
*Open Web Application Security Project (OWASP)*

**Abstract:** SAML assertions are becoming popular method for passing authentication and authorisation information between identity providers and consumers using various single sign-on protocols. However their practical security strongly depends on correct implementation, especially on the consumer side. Somorovsky and others[1] have demonstrated a number of XML signature related vulnerabilities in SAML assertion validation frameworks. This article demonstrates how bad library documentation and examples can lead to vulnerable consumer code and how this can be avoided.

## SAML and SAML vulnerabilities

The article by Somorovsky and others describes a number of vulnerabilities caused in most cases by an incorrect implementation of generally secure SAML model. SAML language is used to construct authorisation statements (*assertions*), whose authenticity is protected by XML digital signature applied over the assertion.

For example, the following XML structure (an identity provider response containing an embedded and signed SAML assertion) could contain a statement such as *"Alice is authorised to use Secret service"*. The assertion has an identifier (1) and the digital signature refers to that identifier.

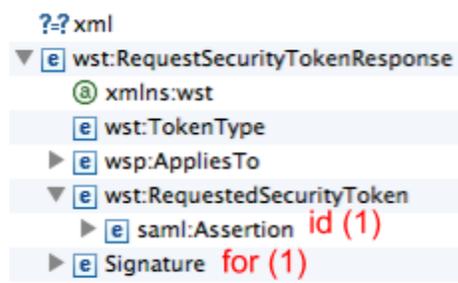

Most of the attacks demonstrated in Somorovsky and others' paper are possible because vulnerable applications make a number of assumptions, for example that the token will be always properly formed XML document, compliant with the SAML schema and with properly linked references.

One of the attacks described in the paper is executed by constructing a false SAML token with structure demonstrated on the below picture. The original assertion (1) is moved to a wrapper node and new, malicious assertion (2) is added (saying, for example, that it's no longer Alice, but Eve who is authorised to use the service).

---

[1] "On Breaking SAML: Be Whoever You Want to Be", Juraj Somorovsky, Andreas Mayer, Jorg Schwenk, Marco Kampmann, Meiko Jensen, 2012

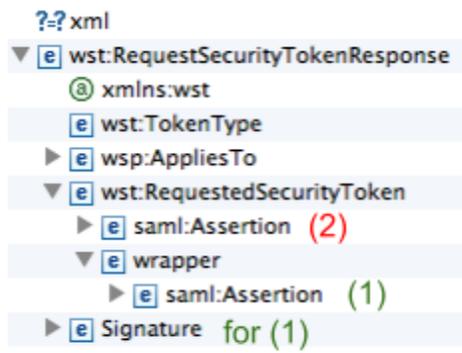

Because the original assertion's body was not modified, the document will still pass digital signature validation. What happens next depends fully on assumptions made by the programmers.

**Opportunistic programmer versus code examples**

Programming XML document processing in Java is non-trivial due to complexity of the programming interface especially if techniques such as namespace resolution and keystore management are involved. For this reason, most programmers given this task would immediately resort to tutorials, examples given in vendor documentation and online discussion forums such as StackOverflow.

> **Popular documentation, forums and tutorials frequently contribute to replication of vulnerable design, either because they were intended for brevity and readability, or because their authors were unaware of possible security issues.**

Most examples on how to select an XML element would use getElementsByTagName method, which - under Java API - is just the simplest way of doing this. Sample from a 2012 bug report for JIRA:

```
NodeList nodes = doc.getElementsByTagName("saml:Assertion");
```

The function returns **a list of all requested elements** in the document. A straightforward intuition is that in a *properly formed* document there will be just a single `Assertion` tag, which explains why the author picked the first item here:

```
element = (Element) nodes.item(0);
```

Unfortunately this approach is exactly what makes the SAML wrapping attacks possible. The attacker exploits the "single and first" tag assumption by placing a new, malicious assertion before the original one and this is never detected, because at the same time most of these examples work on XML documents that are **not validated**.

Oracle documentation uses similar getElementsByTagNameNS method in its XML digital signature validation example:

```
NodeList nl = doc.getElementsByTagNameNS(XMLSignature.XMLNS, "Signature");
```

There is a basic sanity check to handle badly formed input — whether the document has **any** signature elements at all:

```
if (nl.getLength() == 0) {
      throw new Exception("Cannot find Signature element");
}
```

But presence of superfluous signature elements is not checked and the example goes on to pick the first element from the list for further processing:

```
DOMValidateContext valContext = new DOMValidateContext(new KeyValueKeySelector(),
nl.item(0));
```

Assumptions made by authors in both examples will result in assertion and signature wrapping, respectively, as described in Somorovsky paper.

## Secure validation of SAML assertions

*Note: all code samples given below are taken from reference implementation [java-saml-validator](java-saml-validator).*

SAML document validation consists of the following steps:

1. Parsing the XML document, which includes structure validation based on supplied schema;
2. Digital signature validation, which verified authenticity and integrity of the assertion embedded in SAML document.

The first step, schema validation, might prevent XML manipulation attacks such as wrapping (it will not if schema contains "any" extensions, see below). The second step, signature validation, prevents forgery.

Each of these steps has to be successful for the whole validation to complete.

**Recommendation:**

- **Always perform schema validation on the XML document** prior to using it for any security-related purposes.

### Schema compliance validation

XML validation process is performed by the default XML parser supplied as part of standard Java library. Validation has to be explicitly enabled as it's not on by default (see below) and proper schema documents need to be supplied. The first step is to initialize document builder factory:

```
DocumentBuilderFactory factory = DocumentBuilderFactory.newInstance();
```

SAML tokens are complex XML documents with a lot of external references, so the factory has to be put into namespace mode:

```
factory.setNamespaceAware(true);
```

The parser must be put into **validating mode**, which is the most important step of the process:

```
factory.setValidating(true);
```

The next step is to supply the parser with the schema of the validated SAML token. This is also key step, as all validation will be performed against the schema and invalid or non-existent schema will result in failed validation.

XML schema security

The parser must be supplied with the schema of the validated document, which is passed in `schemaFile` parameter below as file path:

```
factory.setAttribute("http://java.sun.com/xml/jaxp/properties/schemaLanguage",
    XMLConstants.W3C_XML_SCHEMA_NS_URI);
factory.setAttribute("http://java.sun.com/xml/jaxp/properties/schemaSource",
    new InputSource(schemaFile));
```

Schema specifies allowed structure of the XML document and the validation is performed against the schema statements.

> **Strength of validation depends fully on schema begin precise in describing the intended document's structure. Schemas may be written using very relaxed, wildcard statements[2] in which case malicious structures may still get through validation stage.**

Because of this, solution architects in high security environments should **verify third party schemas prior to using them for validation**, even if they come from standardisation bodies. Schemas should be manually edited and wildcard statements can be removed ("schema hardening"), as described in RUB paper from 2013[3].

In addition, most XML parsers will use schema namespaces expressed as URLs to automatically download missing schemas on the run. This has negative impact on both performance (as downloads are not cached by default) and security (as schema location may be tampered with).

Here is an example of an namespace identifier that is schema address at the same time, and XML parser will use it if download is successful:

```
<s:Envelope xmlns:s="http://www.w3.org/2003/05/soap-envelope">
```

This behaviour should be constrained using parser configuration — in this case limiting approved schema locations to local files and JAR:

---

[2] These are `xs:any, processContents="lax"` (or `"skip"`) and `namespace="##any"` (or `"##other"`), according to XSpRES article (2012).

[3] Meiko Jensen, Christopher Meyer, Juraj Somorovsky, and Jorg Schwenk, "On the Effectiveness of XML Schema Validation for Countering XML Signature Wrapping Attacks", 2013

```
factory.setAttribute(XMLConstants.ACCESS_EXTERNAL_DTD, "file,jar");
factory.setAttribute(XMLConstants.ACCESS_EXTERNAL_SCHEMA, "file,jar");
```

The SAX[4] platform introduced a standard interface EntityResolver[5] that is called by the XML parser to supply all required schemas on demand. By using the resolver the application retains full control over the schemas that are returned and allows returning only schemas that come from a trusted source, were inspected and, possibly, hardened.

The factory can take one more recommended argument that would limit possibility of a denial of service attack by resource exhaustion:

```
factory.setFeature(XMLConstants.FEATURE_SECURE_PROCESSING, true);
```

Depending on the implementation of the XML parser there may be additional features available, some of which are security related. See Xerces features for description and reference source code for examples.

Finally an actual parser is derived from the factory object:

```
DocumentBuilder db = factory.newDocumentBuilder();
```

Summary:
- **Always use local, trusted copies of schemas for validation.**
- **Never allow automatic download of schemas from third party locations.**
- If possible, **inspect schemas and perform schema hardening**, to disable possible wildcard-type or relaxed processing statements.

Entity resolver

As mentioned above, the XML parser can by default download external schemas but this is undesirable. When automation is disabled (see above), EntityResolver class needs to be implemented to deliver schemas to the parser on demand:

```
SamlEntityResolver res = new SamlEntityResolver();
db.setEntityResolver(res);
```

Full implementation of the EntityResolver class is available on GitHub. The class is called on demand by the parser and takes two arguments: *publicId* and *systemId*. The latter is the namespace identifier that XML parser finds in parsed elements. The task of the resolver is to return schema or DTD contents for this particular URL. Sample extract from the implementation:

```
if (systemId.endsWith("oasis-200401-wss-wssecurity-secext-1.0.xsd")) {
    file = "schemas/oasis-200401-wss-wssecurity-secext-1.0.xsd";
}
```

Error handler

---

[4] Simple API for XML (SAX) 1.0
[5] Interface EntityResolver, since SAX 1.0

A validating parser should also use a custom error handler, which will be called by the parser for any validation errors. Source code of the SamlErrorHandler class will not be shown here, as it basically only outputs an message for every error and it's available on GitHub.

```
SamlErrorHandler err = new SamlErrorHandler();
db.setErrorHandler(err);
```

Now the parser is ready to validate the token:

```
Document doc = db.parse(input);
```

The parse method will load, parse and validate the document for schema compliance and throw exception if the document is not valid.

### XML digital signature validation

The process of schema validation confirms correct structure of the SAML token, but not authenticity and integrity of the embedded assertion. For this purpose the digital signature needs to be validated against a trusted public key of the signer.

W3C XML Signature Syntax and Processing describes the process as "Core Validation", referring to step 3 as "Reference Validation" and step 2 as "Signature Validation". Signer's certificate validation using CRL or OCSP is not described in this standard. The process is described with slightly more details in XML Signature Syntax and Processing Version 2 and "establishing trust" in the signer's key is explicitly listed as a requirement. Some further guidance is provided in XML Signature Best Practices.

Technically, digital signature validation itself is a complex process that should involve the following steps:

1. Establishing trust for signer's public key certificate via trust anchor (CA certificate), trust path validation, revocation and purpose checks[6]. This is traditional PKI model, implemented in most SSL implementations and required in heterogenous environment.
2. Cryptographic validation of document signature's authenticity with signer's public key to confirm signature's authenticity.
3. Cryptographic validation of signed hash of the document against canonicalized hash computed from received document to confirm document's integrity.

Technically, the first step is most complex as it requires validation of certificate features such as keyUsage, time constraints and revocation status (using OCSP or CRL), repeated for all certificates in trust path. This is how it works in typical SSL usage scenarios in highly heterogenous environment of SSL servers.

#### Establishing trust

As said before, the digital signature needs to be validated against a **trusted** public key of the signer. The word "trusted" is critical here. The signed document will frequently contain public keys. But as the whole purpose of signature validation is to validate a document coming from an untrusted source, any keys embedded in the document must not be trusted. **They must not be directly used for validation until they get validated themselves.** Unfortunately, available documentation also fails to explain this in clear and explicit way.

---

[6] RFC 5280, "Internet X.509 Public Key Infrastructure Certificate - and Certificate Revocation List (CRL) Profile", IETF, 2008

In case of SAML based single sign-on we will usually deal with single identity provider authenticated by a single certificate, which greatly simplifies the validation process. We can validate directly against this certificate that we trust explicitly, without the need for full path validation via trust anchor.

Signature validation using [XML Digital Signature API](#) expects a "key selector" function that will provide the validation engine with proper key for validation, as described below.

### Explicit trust

In the simplest scenario the architect would feed a static, trusted public key certificate to the validation function. They key can be obtained using trusted channel from the identity provider and stored in application's configuration. This is the option that we have selected for this article:

```
XMLSignatureFactory fac = XMLSignatureFactory.getInstance("DOM");
DOMValidateContext valContext = new DOMValidateContext(new
    StaticKeySelector(keyFile), signature_element);
```

The `StaticKeySelector` will return public key component of the X.509 certificate stored in DER format in file whose name is passed in keyFile parameter (it's full source code is available on [GitHub](#)). This is the simplest method, but other scenarios are possible, as described below..

**Summary:**
- **If you only expect only one signing key, use `StaticKeySelector`.** Obtain the key directly from the identity provider, store it in local file and ignore any `KeyInfo` elements in the document.

### KeyInfo, KeyValue and X509Certificate elements

As mentioned above, XML signed documents will frequently have a hint element `KeyInfo` that contains signer's certificate, public key or just its identifier. While this may be required in heterogenous environments with many identity providers, **authenticity of these keys must be validated before they can be used for document validation.**

[XML Digital Signature API](#) documentation for Java SE 6 provides example of key selector ([KeyValueKeySelector](#)) that impressionably consumes the key provided in `KeyInfo` structure without any validation. The article briefly mentions its insecurity but offers no secure alternative.

A newer Java SE 7 tutorial ([Programming With the Java XML Digital Signature API](#), code sample 8, class `X509KeySelector`) makes it even clearer and refers to [Java PKI Programmer's Guide](#) but still does not link to any secure implementation.

Much better implementation is available as part of Oracle examples collection in [X509KeySelector](#) class. It has the same name as the above class, but it implements logic based on explicit trust. The trusted key is expected to be stored in local key store ([JKS](#)), which can be created with [keytool](#) program. The class would look up the `KeyInfo` element and if it finds the certificate in local JKS, it would treat it as trusted for validation.

**Summary:**
- Again, **if you only expect only one signing key, use `StaticKeySelector`.** Avoid the complexity of key stores and key selectors if possible. See previous section for details.
- **If you expect more than one signing key, use [X509KeySelector, the JKS variant](#).** Obtain these keys directly form the identity providers, store them in local JKS and ignore any KeyInfo elements in the document.
- **If you expect a heterogenous signed documents** (many certificates from many identity providers, multi-level validation paths), implement full trust establishment model based on [PKIX](#) and trusted root certificates.

Signature and Assertion elements

Digital signature in XML document is stored in `Signature` element. The data whose authenticity is verified is stored in `Assertion` element[7] (namespaces omitted for clarity). Both these elements must be known to the validation function.

As explained before in the context of Oracle tutorial **using getElementsByTagName for that purpose is a bad practice in a document that was not validated against schema** and facilitates signature and assertion wrapping attacks.

Theoretically, after successful schema validation[8] it should be not likely that the document will contain structure anomalies such as wrapping, but, as mentioned above, publicly available schemas frequently contain relaxed syntax. So while validation limits likelihood of attack, it does **not** guarantee that they won't be present in validated document. Because of this **absolute** XPath should be used even on validated documents.

**Summary:**

- **Always validate received XML document against schema** prior to any security related validations.
- **Never used getElementsByTagName to select security related elements** in an XML document without prior validation.
- **Always use absolute XPath expressions to select elements**, unless a hardened schema is used for validation.

Using XPath to select elements

---

[7] This article uses `Assertion` as example, but it depends only on document structure.
[8] Note that this is only guaranteed if validated against a hardened schema, not containing "any" extensions.

XPath allows absolute and unambiguous addressing of elements in XML document, which should be the preferred way to select them for security validation purposes. As discussed above, even schema validated document may still contains wrapping attacks and using absolute XPath further limits feasibility of the attack.

For example Web Services Trust Language (WS-Trust) specification would give the following addresses for these key elements in WST tokens, respectively:

- `/wst:RequestSecurityTokenResponse/ds:Signature`
- `/wst:RequestSecurityTokenResponse/wst:RequestedSecurityToken/saml:Assertion`

Samples provided with this article use the following XPath addresses:

- `/soape:Envelope/soape:Body`
- `/soape:Envelope/soape:Header/wsse:Security/ds:Signature`

Further protection can be provided by XPath hardening as proposed in 2012 paper on XSpRES,[9] a robust XML signature validation interface. The hardening replaces namespace-based references to elements with very precise references to a specific namespace and element. The above `/soape:Envelope/soape:Body` address would be replaced by the following hardened address (single line, wrapped for readability):

```
/*[local-name()="Envelope" and namespace-uri()="http://schemas.xmlsoap.org/soap/envelope/"][1]/*[local-name()="Body" and namespace-uri()="http://schemas.xmlsoap.org/soap/envelope/"][1]
```

The reference code accompanying this article provides automatic conversion between user-supplied XPath and the hardened form used internally.

To run XPath evaluation we need to first initialize the factory and supply a method that will resolve abbreviated namespaces to their full identifiers (for example, "ds" in "ds:Signature" will be resolved to full URL for this schema). Full source code of `SamlNamespaceResolver` is available on GitHub.

```
XPath xpath = XPathFactory.newInstance().newXPath();
xpath.setNamespaceContext(new SamlNamespaceResolver(doc));
```

`Signature` element can be now extracted:

```
Element signature_element = (Element)
xpath.evaluate("/wst:RequestSecurityTokenResponse/ds:Signature", doc,
    PathConstants.NODE);
```

As well as `Assertion` element:

```
Element assertion_element = (Element)
xpath.evaluate("/wst:RequestSecurityTokenResponse/wst:RequestedSecurityToken/saml:
```

---

[9] Christian Mainka, Meiko Jensen, Luigi Lo Iacono, Jörg Schwenk, "XSpRES: Robust and Effective XML Signatures for Web Services", 2012

Assertion", doc, XPathConstants.NODE);

Signature validation follows with initialising signature factory object:

`XMLSignatureFactory fac = XMLSignatureFactory.getInstance("DOM");`

In this step `StaticKeySelector`, as discussed above, is referenced. The argument is file path to the single, trusted X.509 certificate:

```
DOMValidateContext valContext = new DOMValidateContext(new
    StaticKeySelector(keyFile), signature_element);
```

Finally, actual signature validation is performed:

```
XMLSignature signature = fac.unmarshalXMLSignature(valContext);
boolean coreValidity = signature.validate(valContext);
```

After positive response was returned from the validation method, the document and assertion are likely to be authentic.

The validated assertion node should be then passed to the business logic, which implements *"see-what-is-signed"* logic as described in Somorovsky's article ("On Breaking…", 2012).

## Reference source code

A reference source code has been published that attempts to implement most of the above recommendations. The code is published openly on GitHub[10]. The code should not be seen as production-ready implementation but rather an attempt to create a reference implementation in Java.

Its usage is demonstrated in the attached JUnit test suite. First validator object is created:

```
Validator val = new Validator("documents/signer1.der",
    "schemas/soap-envelope.xsd",
    "/soape:Envelope/soape:Header/wsse:Security/ds:Signature",
    "/soape:Envelope/soape:Body");
```

Its positional parameters are as follows:

- signer's X.509 certificate in DER encoding,
- validated document's schema to validate against,
- XPath expression pointing to the digital signature element,
- XPath expression pointing to the validated element (Body in this case).

In the next step, setIdAttribute method is used to point the validator to Body element's attribute that contains its identifier. It's required both from functional and security point of view:

- SOAP envelope schema does not specify Id field so XML signature validator will throw an

---

[10] https://github.com/kravietz/java-saml-validator

exception if this is not specified.
- Even if it would be specified, some schemas do not specifity it explicitly as ID type, which would have the same result on newer JDK (1.6.21+).
- Manipulation with the ID element can facilitate attacks.

```
final String idNamespace =
"http://docs.oasis-open.org/wss/2004/01/oasis-200401-wss-wssecurity-utility-1.0.xsd";
val.setIdAttribute(, "Id");
```

Finally, the validation is performed that returns true for valid document and false for invalid:

```
boolean result = val.validate("documents/file0.xml");
```

# Conclusions

SAML and, general, XML document validation for security purposes is challenging, partially due to complexity of the available programming interfaces and insufficient awareness of known vulnerabilities. Existing documentation, tutorials and forum discussions frequently present simplified solutions that ignore security aspect and promote vulnerable patterns.

The goal of this study was to analyse most frequent issues and build a reference XML validation template for Java that demonstrates the secure approach.